\documentclass[prl,aps,showpacs]{revtex4}

\usepackage{amsfonts}
\usepackage{amsmath}
\usepackage{amssymb}
\usepackage[dvips]{graphicx}
\usepackage{subfigure}

\newcommand{\bigdot}{\,{\raise.3ex\hbox{\circle*{2}}}}
\newcommand{\smalldot}{\,{\raise.3ex\hbox{\circle*{1}}}}
\newcommand{\<}{\langle}
\renewcommand{\>}{\rangle}
\newcommand{\be}{\begin{equation}}
\newcommand{\ee}{\end{equation}}
\newcommand{\bea}{\begin{eqnarray}}
\newcommand{\eea}{\end{eqnarray}}
\begin{document}

\title{Grover-like  search via Frenkel exciton trapping mechanism}

\author{ A. Thilagam} 
\email{thilaphys@gmail.com}
\affiliation{Applied Centre for Structural and Synchrotron Studies, University of South  Australia, Mawson Lakes,
SA  5095, Australia}
\date{\today}
\begin{abstract}
We propose the physical implementation of a Grover-like search problem
 by means of Frenkel exciton trapping  at
a shallow isotopic impurity against a background of
 competing  mechanisms.
The search culminating at the impurity molecule,
designated the ``winner" site, 
 is marked by its enhanced  interaction with 
acoustic phonons at low temperatures. 
The quantum search  proceeds with  the assistance of an
 Oracle-like exciton-phonon interaction  which 
addresses only the impurity site,
via the Dyson propagator within the Green's function formalism.
The optimum  parameters of a 
graph lattice with  long-range intersite interactions 
required to trap the exciton in the fastest time are
determined, and estimates of 
 error rates for  the naphthalene doped organic system are evaluated.
We extend analysis of quantum search to  a 
fluctuating long-range interacting cycle (LRIC) graph lattice
system.

 \end{abstract}
\maketitle
 \section{\label{intro} Introduction}

Quantum search via Grover's algorithm \cite{Grove} 
provides one possible framework by
which combinatorial search can be speeded up considerably over 
those employing classical rules. For a function
$f(x):\{1,\ldots,N\} \to \{0,1\}$ where $  f(x) = 0$ for $x \ne w$
and $  f(x) = 1$ for $x=w$, the  algorithm locates the "winner"  $w$ 
using of order $\sqrt N$ queries for 
spatial dimensions exceeding two \cite{amber}.
The successful bid is generally represented by an oracle $U_w$ with 
$U_w |x \>=(-1)^{f(x)} |x \>$, and is sometimes treated as a black box
which returns the value 1 for a particular  one or a small set of inputs.
The possibility of widening the choices for  quantum searches
via use of quantum random walks  was first
proposed by Farhi and Gutmann \cite{Farhi} who showed that
the time taken to move from one point of a graph vertex
to another is exponentially faster for  a walker following
quantum rules over one who obeys classical physics. 
Unlike their classical counterparts, paths in quantum walks can 
interfere leading to nonexistence of a quantum walker at a 
vertex.

In the continuous-time quantum walk  formulation \cite{Farhi2}, 
a Hamiltonian with oracle-like properties, $H_w = 1-|w\> \<w|$ is employed with
$|w\>$ as the unknown state. Under this formalism,
the ground state of the Hamiltonian is taken as $ |w \>$ while all other 
equivalent but orthogonal states are considered as 
undesired outcomes of the search. Instead
of the counting procedure, the efficiency of the search  is quantified
 by the time taken to locate ground state $ |w \>$.

In this work we attempt implementation of 
Grover's search via the   Frenkel exciton, a bound system of electron and 
hole which propagates from one site to another in a quasi one-dimensional system
of $N$ molecular sites.  Frenkel excitons are also  modeled as 
delocalized excitation over the real crystal space \cite{Dav}, and as
noninteracting fermions \cite{spano}, the exciton is thus an
ideal example of an extended entangled system. 
 We consider the exciton  transfer mechanism as
analogous to  a quantum walker  on a  graph
$G$ where each vertex point represents a lattice site. 
The route of propagation of the exciton from one site to another
is given by the edge joining one vertex  to another with
the number of edges incident on a specific vertex considerably being simplified 
for one dimensional systems.
During  a quantum search on graphs, a particular  vertex is distinguished from 
all others and a successful outcome occurs when this vertex is reached.
We examine the physical implementation of  Grover's search by 
designating an impurity molecule as the ``winner" located at the special vertex.
The impurity site thus mimics the role of the desired
outcome $ |w \>$. The impurity molecule is distinguished from other
 molecules via the assistance of lattice vibrations, so that the difference between
 the energy band with and without the impurity present is accounted
for by phonons. We note the fine balance involved in locating 
the impurity molecule, while lattice vibrations are needed to define
vital properties of the impurity molecule, they should not be strong
enough to hamper the search. 
The analysis can be extended to the existence of more than one
impurity atoms, however we restrict our case to a single ``winner". 

 \section{\label{trap} Exciton trapping mechanism}

The trapping of excitons in molecular crystals has been well 
documented in the literature \cite{Dav}. 
The overlap of intermolecular
wavefunctions is weak in organic molecular crystal systems such as anthracene and 
naphthalene \cite{Dav},   electronic
wavefunctions are thus localized at individual molecular sites.
 The exact form of these functions are not needed to derive salient
properties of Frenkel excitons, however every molecular site 
has an equal probability of being excited by an incident
photon due translation symmetry, a salient feature which  
has important implications for  quantum parallelism.
While a Frenkel exciton can be visualized as being smeared out in space, it 
can also be modeled as a bound electron-hole 
quasiparticle hopping from one site to another.

Trapping sites are categorized as  either impurity
molecules (chemical traps) or dislocated molecules (physical traps), 
with the trap state assumed to be lower than the
 initial exciton state by an energy known as the trap depth. 
The excess energy is absorbed by phonons associated with lattice
vibrations, the trapping mechanism thus occurring as a result of
coupling between the exciton and  phonons. A  transition 
from a delocalized exciton state  to a localized  exciton state occurs
at specific molecular sites. The extend by which neighboring lattice structure is
altered depends on the magnitude of trap depth. 
A shallow  isotopic impurity  participates in the 
dynamics of the delocalized states of the host lattice system, 
unlike a deep impurity state. The
energy state of the  deep impurity   
falls far beyond the energy band states of the 
host molecular system. In this work, we
consider introduction of a  shallow  isotopic impurity which 
alters the crystal symmetry only marginally, but one that eventually traps an exciton in the
final state. We thus assume that the trap depth is present in the range
of phonon bandwidths, so that a single phonon separates the host band
and impurity trap. The trap depth may be viewed as a point of
maximal lattice relaxation, with the excitation moving from an
unrelaxed lattice structure to the relaxed lattice state at the
impurity site, the surplus energy being converted into lattice energy.
The discrete exciton spectrum in a  finite one dimensional crystal system differs
from the continuous  spectrum of the infinite 1D system.
Hence a single discrete level
corresponding to the self-trapped state remains distinct
in the infinite chain unlike the discrete energy levels of  the finite chain.

We note that an initially delocalized
 exciton  examines  all of the sites simultaneously in a superposed form
before reaching its target.   
The exciton lifetime therefore yields a measure of the efficiency of search
from a quantum-information perspective.
The absorption of a photon prior to search triggers the initialization
 process whereby an equal weighted superposition of all states
is created. The exciton  eigenvector appear as
\be
\label{ex}
|{\bf K} \> =  N^{-1/2} {\sum_{\bf l}}
e^{i {\bf K . l}} B_{\bf l}^\dagger |{\bf 0} \> 
\ee
where {\bf K} is reciprocal lattice vector, $|{\bf 0} \> $
is the vacuum state with all molecules in ground state
 and 
$B_{\bf l}^\dagger$ is the creation operator of exciton
at position coordinate  ${\bf l}$. $N$  
is the number  of  unit cells in the crystal which increases
with the number of molecules for each unit cell. 
We note that Eq. (\ref{ex}) is the result of a single
step of a naturally occurring process involving photon absorption that
require the equivalent of $\log_2(N)+1$ qubits to store the superposed state.
In Eq. (\ref{ex}), we have dropped notations associated with the exciton
spin, however we will  consider spin indices in mechanisms involving exciton
formation and annihilation in forthcoming sections of this work.

The  Frenkel exciton creation operator with wavevector ${\bf K}$
can be easily obtained using the Fourier transform of Eq. (\ref{ex})
\be
\label{exF}
B_{\bf K}^\dagger =  N^{-1/2} {\sum_{\bf l}}
e^{i {\bf K . l}} B_{\bf l}^\dagger 
\ee
The exciton creation operator $B_{\bf K}^\dagger$ localized in ${\bf K}$-space
is delocalized in real space \cite{Dav}. The motion of
the exciton in molecular systems is governed by an Hamiltonian
derived using a tight-binding model \cite{Dav,Craig,toy}
\be
\label{exH}
\hat{H}_{ex}= \sum_{\bf l}\left( {\Delta E} +
{\sum_{\bf m \neq l}} D_{\bf l,m} \right )B_{\bf l}^\dagger B_{\bf l}+
\sum_{\bf m \neq l} M_{\bf l,m} B_{\bf l}^\dagger B_{\bf m}
\ee
where $\Delta E$, the on-site (intra-site) excitation energy at equilibrium 
is the same at all sites due to translational symmetry. 
$D_{\bf l,m}$ is the dispersive interaction matrix element
which determines  the   energy difference between a pair of 
excited electron and hole at a molecular site and ground state electrons
at neighboring sites \cite{Dav,Craig}.  $M_{\bf l,m}$  the 
electron transfer matrix element between molecular sites at ${\bf l}$
and ${\bf m}$. Using Eq. (\ref{exF}), we obtain a Hamiltonian 
 diagonal in ${\bf K}$ space
\bea
\label{exHk}
\hat{H}_{ex} &=& \sum_{\bf k} E_0({\bf k}) B_{\bf k}^\dagger B_{\bf k} \\
\label{exenergy}
E_0({\bf k})&=&\Delta E+\sum_{\bf m \neq 0} D_{\bf 0,m}+ M_{\bf 0,m} \exp(i{\bf k.m})
\eea
where $E_0({\bf k})$ is energy of the exciton of wavevector ${\bf k}$ and subscript
$0$ denotes the absence of lattice site fluctuations.
For simple $d-$dimensional lattice with nearest neighbor transfer energy
 $M_{\bf l,m} \sim J$, we obtain the well known dispersion relation
 $E_0({\bf K})=\Delta E+2 J \sum_{i=1}^d \cos(K_i)$ with the exciton band
extending from $\Delta E -B$ to $\Delta E+B$. $B=2 J d$ denotes the half band width
and we emphasize its critical dependence on efficiency of exciton transport 
on designated graph structure of molecular sites. 

\subsection{\label{bwidth} Excitonic bandwidth in 
 graph lattice with  long-range intersite interactions}

The bandwidth $B$ which is determined by the 
energy span  of the host lattice
 over a spectrum of wavevectors,  
 is a vital  property of the exciton.
It  influences  excitonic coherence quantities
such as its delocalization length and relaxation mechanism,
and provides a measure of time needed for the exciton to move
away from its initial site of excitation. It is 
intimately linked with arrangement of 
superpositions of single-site excitations of the 
initial state. In a recent work on photosynthetic organic systems \cite{alex}, it was shown that
the efficiency of excitation transfer  depends both on the symmetry properties
of such superposition states and on the number of sites
among which the excitation is initially delocalized, with results
showing an optimal delocalization length for which  transfer time is a minimum. 
To illustrate the dependence of the trapping time on 
size-effects associated with long-range excitonic coherences
as well as underlying structure of
the lattice points, we consider  
 transfer matrix element $M_{\bf l,m}$ of the form 
\be
\label{lrange}
 M_{\bf l,m} = \frac{J}{|m-l|^\mu}
\ee
where $M_{\bf m,m}=0$ and intersite energy term
 $J > 0$. This form of the matrix element has recently been shown
to reveal various rich and interesting properties for a select range
of $1 < \mu < \frac{3}{2}$ \cite{maly1,maly2,maly3}.
In the absence of lattice vibrations, the
eigenstates of Eq. (\ref{exH}) form waves with momenta $K=\frac{2 \pi k}{N}$ within
the Brillouin zone $k \in [-\frac{N}{2},\frac{N}{2}]$
with  energies
\bea
\label{Energ}
E(K)&=&2 J \sum_{n=1}^{N} \frac{\cos(K n)}{|n|^\mu}\\
\nonumber
&=&\sum_{s=+,-} {\rm Li}_{\mu}(e^{i K s})-\sum_{s=+,-}e^{i (N+1) K s}\Psi(e^{i K s},\mu,N+1)
 \eea
where  ${\rm Li}_{\mu}(a)$ is the Polylogarithm function and $\Psi(z,\mu,a)$ is the
Lerch transcendent function \cite{dalai}.
In the limit where $K \rightarrow 0$ which occurs at the 
upper band edge, we use a simplified dispersion relation obtained as
$E_{K}=E_0(\mu)-J A(\mu) |K|^{\mu-1}$ \cite{maly1} where $E_0(\mu)$
is the upper band-edge energy and $A(\mu)$ is a dimensionless constant. 
$E_0(\mu)$ and $A(\mu)$ can be obtained using series expansion of Eq. (\ref{Energ}). 
An approximate but simple relation between the bandwidth
and system size $N$ follows at the upper band edge 
($1 < \mu < \frac{3}{2}$)
\be
\label{bw}
B(K \rightarrow 0, N \rightarrow \infty) \approx E_0(\mu)(1-A(\mu)^\prime N^{1-\mu})
 \ee
where $A(\mu)^\prime < 1 $ is a small dimensionless quantity.
 While the states are extended over the whole crystal structure at  $\mu < \frac{3}{2}$,
it has been shown that 
localization effects appear for higher values of $\mu$ for systems 
with {\it uncorrelated} diagonal disorder and long-range coupling of the form
in Eq. (\ref{lrange}) \cite{maly1}. We note that  lattice vibrations, as considered in our work, 
provide a fabric of {\it correlated} diagonal disorder and therefore localization effects
may be present with properties different from those obtained by Rodr\'iguez et al \cite{maly1}.
Nonetheless, we consider that the system obeying long-range coupling of the 
form in Eq. (\ref{lrange}) still undergoes vital changes at the critical point
when $\mu =\frac{3}{2}$

 \section{\label{time} Estimating search time using Green's functions.}
Implementation of the quantum search process involves
the following crystal Hamiltonian with the impurity at site $w$ 
\bea
\label{exHim}
\hat{H}_{T}&=& \hat{H}_{0}+ \hat{H}_{w}\\ \nonumber
\hat{H}_{0}&=&\hat{H}_{ex} + \sum_{\bf q} \hbar \omega({\bf q})b_{\bf q}^\dagger b_{\bf q}
\\ \label{exwin}
\hat{H}_{w}&=&\Delta p B_{w}^\dagger B_{ w}
 \eea
where $\hat{H}_{ex}$ is given in Eq. (\ref{exHk}) and the second term in $\hat{H}_{0}$ denotes the
phonon energies. $\hat{H}_{0}$ correspond to a impurity-free lattice structure.
$\hat{H}_{w}$ denotes a trapped exciton at site $w$, with trap depth $\Delta p$  defined
as the difference between a state of zero phonon and one where interactions between 
exciton and phonons leading to absorption of a phonon is optimum. 
We consider the presence of just one impurity site and assume that
the  translational symmetry of the molecular system is left intact. This is justified
for isotopic impurities which mimic the molecular structure of the host
molecules quite well. We note that  the lowered
energy $\Delta p$ associated with the Hamiltonian $\hat{H}_{w}$
effectively marks the searched state, $|w\>$.
As discussed in Ref. \cite{childs}, solving the Grover's problem involves
the determination of the lowered energy, and hence 
energy of the Hamiltonian in Eq. (\ref{exHim}). This translates to 
evaluation of the trapping time of an exciton initially located at
a site other than at $|w\>$.

We emphasize that the driving term $\sum_{\bf q} \hbar \omega({\bf q})b_{\bf q}^\dagger b_{\bf q}$
associated with a background of boson bath assist with the Oracle search operation.
A  resonant coupling term associated with 
exciton-phonon interaction at all host sites has been intentionally
left out in Eq. (\ref{exHim}). This is based on  the assumed presence of 
an Oracle-like mechanism   which ensures that all
phonon-assisted relaxation (one-phonon emission and Raman processes) 
cumulates  at the  impurity site, bypassing
other lower energy sites, during the search process. 
 An Oracle is a component of an algorithm 
which can be regarded as a "black box" that 
recognizes solutions to a given problem \cite{deutsch}.
The inner dynamics of the Oracle
is  generally unknown, however at very low temperatures its 
inner workings appears almost apparent. Raising the temperature of the crystal medium
therefore involves the risk of the Oracle malfunctioning, as will be shown 
in detailed evaluations in the next Section. We therefore 
address the impurity trapping problem at very low temperatures (close to $0$K),
a limit in which the mean-free path associated with
 phonon scattering is large
due to a linear scaling of exciton scattering rate with temperature.
The exciton is more likely to be trapped at the
impurity site than to bypass it. Various other
 competing  mechanisms which act to
 disrupt the search and introduce errors will also be
discussed subsequent  to the current Section.

The exciton-phonon coupling Hamiltonian associated with the interaction
between the exciton at site $w$ and phonons can be written as \cite{Dav}
\be
\label{exphon}
\hat{H}_{exp}=B_{w}^\dagger B_{w} \sum_{\bf q} \chi(q) (b_{\bf q}^\dagger +b_{\bf q})
\ee 
where  coupling function $\chi(q)$ arises due to  
acoustic phonons dominant at low temperatures.
$\chi(q)$  is dependent only on the phonon wavevector $q$ 
\bea
\label{coup}
\chi(q)^2&=& \frac{1}{N}\hbar \omega({\bf q}) E_{LR}\\
\nonumber
E_{LR}&=&\frac{E_D^2}{2 I v^2}
 \eea
where phonon frequency $\omega({\bf q})= v |q|$, $v$ being the velocity of sound in the
crystal medium and $E_{LR}$ is the lattice relaxation energy. $I$ is the mass coefficient
of host molecules, taken as almost the same as that of the impurity and deformation
potential $E_D$ remains an invariant throughout all lattice sites. A coupling coefficient
$\gamma = E_{LR}/B$ yields a measure of the scattering regime of the crystal system.
For strong lattice vibrations at high $\gamma$, the exciton is less likely to hop and 
tends to remain localized  at its initial site. At small $\gamma$, the exciton is in a 
delocalized state, until it reaches the site at $|w\>$. Thus the fine interplay between
 $E_{LR}$ and $B$ plays a crucial role in the  efficiency of the quantum search. While
lattice vibrations are essential in  locating the impurity site, these should not be 
too strong to disrupt the intermolecular cooperativity which determines the 
search process. 

We emphasize two critical operations of the Grover's search which is satisfied by
 Eqs. (\ref{exwin}) and (\ref{exphon}). While Eq. (\ref{exwin}) provides a local
operation which  distinguishes
the impurity site from other sites, Eq. (\ref{exphon}) provides an oracle-like
operation which addresses only the impurity site via the evolution mechanism
present in the Green's function formalism. The Green's function for an exciton at time
$t$ is given by \cite{Dav,suna}
\bea
\label{g1}
G({\bf k},t)&=&-i \<0:n_{\bf q}|T\{B_{\bf k}(t) U(t,0)B_{\bf k}^\dagger(0)\}|0:n_{\bf q}\>
\\ \label{ge}
B_{\bf k}(t)&=&B_{\bf k}(0) \exp(i E({\bf k})\; t)
\eea
where $|0:n_{\bf q}\>$ denotes a state with 
$n_{\bf q}$  phonons of  wavevector ${\bf q}$
and zero exciton population. $T$ is the time ordering operator and
$U(t,0)$ is the Dyson function
\be
\label{Dy}
U(t,0)=T \exp\left(-i \int_0^t \exp(i \hat{H}_{T} t') \hat{H}_{exp}
\exp(-i \hat{H}_{T} t') dt'\right)
\ee
where $\hat{H}_{T}$ and $\hat{H}_{exp}$ are given in  Eqs. (\ref{exHim}) and 
(\ref{exphon}).
A direct relation between the Green's function $G(k,t)$ in 
 Eq. (\ref{g1}) and exciton energy  $E({\bf k})$ has been
derived by Craig et al \cite{craigG1} at $t > 0$,
\be
\label{g2}
G({\bf k},t)= -i \exp\left[-i \;E({\bf k})\; t -\<F\>\right]
\ee 
At temperature close to $T=0$K, $\<F\>$ can be  obtained in 
an explicit form  \cite{craigG1,craigG2}
 \bea
\label{g3}
\<F\>&=&\frac{1}{N}\sum_{\bf q} |\chi(q)|^2 \left[-\frac{i t}{\eta_{\bf k,q}}+
\frac{1}{\eta_{\bf k,q}^2}(1-\exp(i \eta_{\bf k,q} t) \right ]\\
\label{g4}
\eta_{\bf k,q}&=&\hbar \omega({\bf q})+E_m-E({\bf k})-\Delta p
 \eea
 $E_m$ is the mean energy of the exciton band,
 $N$ is the number of sites and the pre-factor $\frac{1}{N}$
is due to the search for just one winner. 

The imaginary component of the Green's function in 
 Eq. (\ref{g2}) yields
damping rate of the exciton transfer process among lattice sites.
Here we use the damping rate to estimate the time needed to search for
the impurity molecule, $T_s$ 
\be
\label{stime}
\frac{1}{T_s} =\frac{1}{N}\sum_{\bf q} |\chi(q)|^2 \delta(\eta_{k,q})
\ee
In order to obtain an explicit result for $T_s$, we assume that the phonon
modes obey the Debye distribution with  density of states given by $\rho(\hbar \omega) =
3 N (\hbar \omega)^2/(\hbar \omega_D)^3$ where $\omega_D$ is the Debye cutoff
frequency. Using Eqs. (\ref{bw}),  (\ref{coup}) and  (\ref{stime}), we obtain
an approximate expression for search time $T_s$ at $1 < \mu < \frac{3}{2}$
\bea
\label{stimef}
\frac{1}{T_s}  &=& \frac{1}{T_0}+ \frac{1}{T_N} \\ \label{stimef2}
T_0  &\approx&   \frac{\hbar \; E_0(\mu) }{3 \pi E_{LR}}
\left( \frac{\hbar \omega_D^2}{\Delta p^3}\right)\\ \label{stimef3}
T_N &\approx& \frac{T_0} {A(\mu)^\prime} N^{\mu-1}
 \eea
where $E_0(\mu)$ is the upper band-edge energy in  Eq. (\ref{bw}).
In order to highlight the  explicit dependence of $T_s$ on $N$,
we have approximated energy of a single phonon, $\hbar \omega({\bf q}) \approx \Delta p$
instead of using the energy conservation relation in  Eq. (\ref{g4}).
The search time constitutes a $N$-independent  term associated with the maximum 
bandwidth and dependent on the topological parameter $\mu$ 
(see Eq. (\ref{bw})), and 
 a second Grover-like  $N$ dependent term 
which scales as $ N^{\mu-1}$. Thus two concurrent mode of search mechanisms
appears to be involved in Eq. (\ref{stimef}), one associated with 
a quantum coherence effect that is independent of $N$ and another term which
has the well-known  Grover-like  search features  \cite{Grove}.
 At $N \rightarrow \infty$, we obtain a constant
search time $T_s = T_0$ for the system. We note that the maximum
bandwidth associated with  $E_0(\mu)$ in Eq. (\ref{bw}) corresponds to a 
delocalized state of optimum coherence when the search is conducted from all lattice
sites. At the peak of the energy band, the exciton can be considered to exist at all sites
at the same time, in a maximally entangled state of existence. 
The search time $T_0$ associated with the maximum
bandwidth yields the fastest time and consequently the most efficient route
to search the impurity site. For very large
$T_N \ge T_0, \; T_s \sim T_0$, the search route is dominated by the
first coherent term in   Eq. (\ref{stimef}). 

 At the marginal $\mu=\frac{3}{2}$, 
we obtain a term that  equivalent
to the well-known Grover's search term, $\propto N^{\frac{1}{2}}$  \cite{Grove}.
The Grover's algorithm result 
gives the maximal possible probability of finding the desired element \cite{zalka}.
However it only  describes a specific mode to search a list of $N$ sites and
in this regard, the speedup features
in the excess search term $T_N$ of Eq. (\ref{stimef}) at  $\mu < \frac{3}{2}$ is
associated with the peculiarities of the level spacing (of order $\Delta E \sim
N^{-\mu+1}$) at the top band as discussed in Ref. \cite{maly1}. This implies that
the search proceeds with a decreasing number of target sites 
as the energy of the quasi-particle decreases. 
The explicit relation in  Eq. (\ref{stimef}) was obtained by
 utilizing a background boson field coupled with 
 the evolution mechanism of 
the Dyson propagator (see Eq. (\ref{Dy})) within the Green's function formalism.
Such an  approach may facilitate search procedures with
a number of channels for the delocalized excitation  to pursue
 the preferred impurity site efficiently.
The range of $\mu$ considered here provide 
ideal conditions  required to trap the exciton in 
the fastest time, when the quantum search is considered 
accomplished. 
In contrast to the dynamics at the top of the
exciton band, it was shown that the energy level spacing scales as
$\Delta E \sim N^2$ at the bottom  band \cite{maly1}. This means that the search for a
target site located at a higher level would  experience
a slowdown effect as is obvious when substituting higher $\mu$ values in Eq. (\ref{stimef}).
The parallels between the quantum search mechanism and 
physical processes considered so far  is shown in Table 1.
\bigskip
\begin{center}
{{\bf Table 1} \; {\it Parallels  between exciton trapping mechanism and the 
proposed Grover-like search problem}}
\bigskip

\begin{tabular}{|c|c|c|} \hline
{\bf Physical State/Process} & {\bf Implementation} \\ \hline
 Ideal crystal in Vacuum state & Null state $|0\>^{\otimes N}$ \\ \hline
Photon Absorption   & Hadamard-like transformation \\ \hline
 Delocalized exciton & Entangled state\\ \hline
Isotopic impurity molecule & Grover's Target state (``Winner") \\ \hline
Total number of molecules & System size \\ \hline
Phonon field & Marks Target state \\ \hline
Exciton-phonon interaction operator & Oracle-like action \\ \hline
Dyson propagator& Assists Quantum Search  \\ \hline
Expt.  measurement(e.g. Fluorescence) & Search Ends \\ \hline
Annihilation or relaxation processes & Sources of Error \\ \hline
Quantum Zeno Effect & Error Minimization \\ \hline
\end{tabular}
\end{center}
\bigskip

\subsection{\label{ora} Organic molecular  systems with $\mu=3$.}
 
It is important to note that the   marginal exponent  $\mu=\frac{3}{2}$
is not expected in any known molecular system, nevertheless
in organic systems such as Frenkel excitons in 
Langmuir-Blodgett films, molecules experience dipole-type exchange mechanisms
with $\mu=3$ \cite{Dav,Craig}.  For such systems, the exciton bandwidth expression
in Eq. (\ref{bw}) is modified as \cite{maly2}
\be
\label{bworg}
B(K) \approx E_0(\mu)(1-A(\mu)^\prime N^{-2})
 \ee
for ($\mu > 3$).
Using Eqs. (\ref{bworg}),  (\ref{coup}) and  (\ref{stime}), we obtain
\bea
\label{stimeorg}
\frac{1}{T_s}  &\approx& \frac{1}{T_0}+ \frac{1}{T_N} \\ \nonumber
\\ \label{timeN}
T_N &\approx& \frac{T_0} {A(\mu)^\prime} N^2
 \eea
where $T_0$ is given in Eq. (\ref{stimef}). 
We use crystal parameters for the naphthalene  host crystals (${\rm C_{10}H_8}$)
doped with isotopic impurities used in an  earlier work \cite{thila},
with $\Delta p \approx 50{\rm cm}^{-1}$, 
$\hbar \omega_D \sim E_0(\mu) \approx 90{\rm cm}^{-1}$ and $E_{LR} = 0.004$ eV. For
$N=10^2$ and $\mu=3$, we obtain search times 
$T_0 \approx 0.01$ps, $T_N \approx 10-100$ps and overall
$T_s \approx 0.01$ps. The vast difference in search times, $T_0$ and 
$T_N$ suggests that  the search route in organic systems may be dominated by the
first coherent term in   Eq. (\ref{stimeorg}). 
Delayed fluorescence techniques \cite{expt} may be used to detect the impurity site
and confirm the results obtained here. The emission of fluorescence during 
experimental  measurement can be considered to signal the end of the  quantum search.
Recent advances in spectroscopic techniques \cite{expt1}  may  be used to
verify the  femtosecond  time scale  estimates 
predicted for  $T_0$  in 
naphthalene doped organic systems.
Studies of exceptionally efficient
 excitation transfer in photosynthetic systems  has also 
 been explored in recent works \cite{alex}.
In this regard, there is possible application of 
the coherent size-independent, connectivity dependent
   time component $T_0$ derived in Eq. (\ref{stimef}) to studies related to
excitation transfer in   photosynthetic systems and associated phenomenon of quantum
photosynthesis.

\section{\label{error} Processes which compete with Grover-like search process}

In this section, we  consider various  sources of decoherence-type 
mechanisms that will result in improper search
patterns of the impurity site. 

\subsection{\label{errorf}Downward relaxation and decoherence of superposed exciton states}

Exciton-phonon scattering can interfere with the search process by causing 
downward relaxation of states which occupy the upper exciton band levels. The 
possibility of such a process  can be minimized by
 ensuring  that the exciton remains delocalized and hence entangled
over a   number of molecular sites with size equivalence  to the overall number of 
sites to be searched. In the presence of 
decoherence effects due to lattice vibrations, it is thus imperative 
that the coherent length associated with the  superposed exciton state in Eq. (\ref{ex})
should at least exceed the number of sites to be searched. This means that the
exciton state should retain its coherent form at least for the time needed  for the exciton 
to reach the target impurity site, a condition best described by the relation 
\be
T_N < T_{\rm scat}
\ee
where $ T_{\rm scat}$ is the scattering time. Using perturbation theories, the scattering
rate can be estimated using $T_{\rm scat} \sim \; E_{LR} k_B T \rho(E)$ where 
$\rho(E)$ is the normalized density of states of the unperturbed lattice system\cite{toy}. 
For s constant density within the band $(\Delta E-B,\Delta E+B)$, 
$\rho(E) =  \frac{1}{\sqrt{B^2-E^2}} \sim \frac{1}{2B}$ at $\mu=3$.
The density of states, $\rho(E)$ is highly sensitive 
to the parameter $\mu$ which appears in Eq. (\ref{lrange}) \cite{maly2}.
For $\mu=3$, we obtain
\be
\label{scatTime}
T_{\rm scat} \approx \frac{\hbar B}{E_{LR} k_B T}
\ee
where  $E_{LR}$  the lattice relaxation energy is defined in  Eq. (\ref{coup})
and $B$ denotes the half band width. Comparing Eq. (\ref{scatTime})
with Eqs. (\ref{stimef2}) and (\ref{stimef3}), and substituting
the upper band-edge energy  $E_0(\mu) = 2 B$ and $A(\mu)^\prime \approx 1$,
we obtain an explicit condition required for the Grover-like search to proceed successfully
without being affected greatly by downward relaxation 
\be
\label{condrelax}
 N^2 < \frac{3 \pi \Delta p^3}{ k_B T \hbar \omega_D^2} 
\ee
The number of sites that can be searched without disruption increases
at low temperatures and at higher trap depth, $\Delta p$. We bear in mind, however the
possibility that the impurity energy band may become disengaged from the exciton
band with significant loss in translational symmetry due to deep traps associated 
with very large $\Delta p$. 

Another undesirable effect associated with exciton-phonon scattering
is the introduction of composite exciton-phonon characteristics 
to the exciton bandstructure. It can be
seen  that a fixed sum of exciton and phonon wavevectors given by 
 ${\bf K}= \sum_{\bf k,q} {(\bf k+q)}$,
involves  an infinite number of phonons with wavevector ${\bf q}$. This
leads to a weakening effect of an impurity trap due to interference. For instance, the presence
of an isotopic impurity displaces the energy band by $E=\sqrt{B^2+\Delta p^2}$, so that
at bandwidth $B=300{\rm cm}^{-1}$ and  $\Delta p \approx 100{\rm cm}^{-1}$ we obtain
an energy shift of about $\Delta E=16{\rm cm}^{-1}$ below the band-edge. At temperatures
higher than $30$K, it is not uncommon for appearance of band spectra features 
of the order $\sim 10{\rm cm}^{-1}$ \cite{Dav}, this implies  that exciton-phonon coupling 
leads to comparable energy shifts as the impurity state. Therefore
exciton-phonon relaxation may produce ``defective" winners. Hence operating at the 
right temperatures holds the key to maintaining a fine balance between 
 undesirable features of phonons and
its  facilitating  role in energy transport. The calculations above that 
 at temperatures close to $5$K, we expect a clear distinction between ``defective"  and
``genuine" winners when the exciton are trapped at the impurity states displaced from the
exciton band. Another  side effect of 
$T_{\rm scat}$  is manifested in the the phase coherence in Eq. (\ref{exF}) which 
becomes increasingly 
 damped at higher temperatures
\be
\label{exDamp}
B_{\bf K}^\dagger =  N^{-1/2} {\sum_{\bf l}}
e^{i {\bf (K+i k') . l}} B_{\bf l}^\dagger 
= N^{-1/2} {\sum_{\bf l}}
e^{i {\bf K.l}} e^{- {\bf k'.l}} B_{\bf l}^\dagger
\ee
It is well known that the  imaginary component $k'$,
 which appears as a result of the scattering mechanism
in Eq. (\ref{scatTime}), is a source of decoherence in  
  the entangled exciton state.

For the naphthalene  host crystals (${\rm C_{10}H_8}$), 
typical value of $B \sim 100{\rm cm}^{-1}$, and
using $k_B T=3.7$meV at $T=30$K, we obtain $T_{\rm scat} \approx
50$ps which reduces to $500$ps at a lower $T=5$K. These times
are comparable to the search time $T_N \approx 100$ps obtained
earlier,  suggesting  that  downward relaxation mechanism may pose
as a  serious drawback to the Grover's search mechanism at increasing temperatures
($T>10$K). It should noted that $T_0$ (see Eq. (\ref{stimef2}))
 associated with the maximum bandwidth and with coherence 
properties still remains the dominant pathway to energy transfer due to 
the very fast times $T_0 \approx 0.01$ps computed earlier for 
naphthalene  crystal system in comparison
 to  the Grover's search term $T_N$ 
(see Eq. (\ref{stimef3})) as  well as $T_{\rm scat}$.

\subsection{\label{errorf}Exciton formation at impurity site}
 It is possible that an exciton is created
at an impurity site due to the binding  of free charge carriers instead of
trapping of an exciton participating in the 
Grover-like  search  mechanism. In this section, we compute the rates
associated with such a process which acts to disrupt the search process.
In Frenkel excitons, both 
charge carriers are generally created via optical excitation  
on the same molecule. The 
 transition from the initial state of charge carriers to a final
state of trapped exciton via acoustic phonon occurs as
\be
{ \bf e} \;\; + \;\; {\bf h}
\;\;{\longrightarrow }
 \;\;{{\bf  e \atop \vert} \atop {\bf h}}\; + \;  \hbar \omega,
\ee
where ${\bf e, h}$ and  ${\bf e \atop \vert} \atop {\bf h} $ denote  a free electron,
hole and exciton respectively, and $ \hbar \omega$ is
the frequency of an acoustic phonon. 
The initial and final states appear as
\bea
\label{cons1}
|i\> &=&  { \sum _ { \sigma _e , \sigma _h }}
a_{\bf k}^\dagger( \sigma_e) d_{\bf K- k}^\dagger
( \sigma_h) |0; n \> \\ \nonumber
|f\> &=&  B_{{\bf K} - {\bf q} }^ \dagger(\sigma_S)
b_{\bf q }^\dagger |0; n\> 
\eea
where $\sigma _e (\sigma _h)$ denotes the electron (hole) spin, while
the lowest internal exciton state is assumed with spin 
$\sigma_S=0(1)$ for the singlet (triplet) state \cite{thilama}.
$a_{\bf k}^\dagger (d_{\bf k}^\dagger)$ denote the electron (hole)
creation operator of wavevector ${\bf k}$ in their respective energy bands.
The exciton wavevector can be written as a linear combination of all
eigenvectors of the free charge carriers (see Eq. 3 in Ref.\cite{thilama}).
The initial and final state  eigenvalues are obtained using
\bea
\label{cons2}
E_i&=& W_0 + E_g + { \hbar ^2 {\bf k}^2 \over 2 m_e^\ast } + {\hbar ^2
( {\bf K } - {\bf k})^2 \over 2 m_h^\ast }
 + \Gamma_{ph} \\ \nonumber 
E_f&=& W_0 + E_g +\Delta p +  \frac{\hbar ^2 ({\bf K} -{\bf q})^2}
{2 M^\ast} + \hbar \omega ({\bf q }) + \Gamma_{ph} 
\eea
where  $\Gamma_{ph}$ represents the phonon energy that remains unchanged
during the transition and $W_0$ is the total electronic energy of crystal
with all molecule in their ground states.  $m_e^\ast, m_h^\ast$ and $M^\ast$ are the respective
effective masses of the electron, hole and exciton.
The rate of formation of the exciton at the impurity site is evaluated
using the following exciton-impurity-phonon interaction operator 
\be
\label{exphonimp}
\hat{H}_{imp}=\Delta p N^{-1} \sum_{\bf k,k',q} \left[ \frac{\hbar}
{2 I_p \omega ({\bf q })} \right ]^{1/2} ({\bf k-k'})\cdot \hat{e}({\bf q})
\exp[i ({\bf k'-k+q)\cdot p}] \; B_{\bf k}^\dagger B_{\bf k}
(b_{\bf q}^\dagger +b_{\bf q}),
\ee 
where $\hat{e}$ is the unit polarization wavevector and $I_p$ is the 
mass coefficient of the impurity molecule which we assume to the 
same as the mass coefficient of the host molecules. The rate of  formation
of an exciton at the impurity site  is calculated using 
\be
\label{exphonimp}
R_f = \frac{n_i}{\hbar} \sum_{\bf k,q} |\<f|\hat{H}_{imp}|i\>|^2 \; \delta(E_f-E_i)
\ee
where $n_i$ denotes the concentration of free charge carriers.
For the singlet exciton ($\sigma_S=0$), we obtain 
using Eq. (\ref{cons1}),(\ref{cons2}) and (\ref{exphonimp})  an explicit
expression for the exciton formation rate at low temperatures
\be
\label{formrate}
\frac{1}{T_f} = R_f \approx \frac{2 n_i m_e^\ast \; N^\ast \; d_0}
{\hbar^2 \xi v}\left[(E_T-E_b)^2 +\Delta p^2 \right]
\ee
where $N^\ast$ and $d_0$ are the number of molecular sites and 
inter-molecular distance respectively at an edge of the crystal unit cell. 
$E_T$ is the transport gap, the minimum
energy required for the formation of a free electron and
hole pair, $E_b$ is the binding energy, 
$\xi$ is  mass density of the host molecules and $v$ denotes the 
velocity of sound in the crystal. 

For  naphthalene  host crystals (${\rm C_{10}H_8}$)
doped with isotopic impurities, we  use
 $\Delta p \approx 50{\rm cm}^{-1}$, $ (E_T-E_b) =
 3 \times 10^4 {\rm cm}^{-1}$, $n_i \sim 10^{15} {\rm cm}^{-3}$,
$\xi=1283$ kg m$^{-3}$, $v=10^4$cm s$^{-1}$, $N \times d_0= 7.8 \times 10^{-10}$ m,
and obtain  $T_f \approx 10^{-9}$s. We note the crucial dependence
of the formation rate on the carrier concentration, however 
even with a large carrier concentration
$n_i$, a comparatively slow error time is obtained. This can be attributed to 
the  large bandgap in molecular systems, with the requirement that 
charge carriers have to be  optically generated the 
probability of exciton  formation at the impurity site 
is less likely  at low excitation intensities. Moreover due to 
the large binding energies of the Frenkel exciton, participation of 
phonons of relatively higher energies is needed before excitons can be
formed at the impurity site, dissimilar to the situation in semiconductors \cite{thilprb}.
We can therefore conclude that exciton formation at impurity site is not
disruptive to the proposed Grover's like search mechanism, at least in 
naphthalene -doped organic  crystals.

\subsection{\label{errorc}Exciton-exciton annihilation.}
Excitons in close proximity can dissociate due to electrostatic
interactions, creating free charge carriers. The energy 
released from recombination of an electron and a hole
bounded together as an exciton at one molecular site can lead to dissociation 
of exciton at another site, with possible participation of phonons.
Such a process can compete with the proposed Grover-like search  process, hence
 we examine  exciton-exciton annihilation  between 
 two free excitons {\it prior} to exciton trapping at the impurity
site. For simplicity, we consider the two interacting excitons
to be in the singlet spin state ($\sigma_S=0$), the generated hole carrier to be
trapped at a site leaving only the  election free after annihilation. The conservation
of momentum rule ensures that the exciton momentum is transferred to electron
after the annihilation process, in which participation of phonons are
omitted. Due to the small  phonon participation number, the order of magnitude
of the annihilation rates are not greatly affected by  omission of interaction
terms associated with phonons. 
We write the  eigenvectors of the initial and
final states  as
\bea
|i \> &=& B_{\bf K_1}^\dagger B_{\bf K_2}^\dagger|0; n\> \\
|f\> &=& {\sum _{\sigma_e,\sigma_h}} a_{\bf k}^\dagger(\sigma_e) 
d_{\bf k^ \prime}^\dagger(\sigma_h) |0; n\> 
\eea
where $a_{\bf k}^\dagger (d_{\bf k}^\dagger)$  which denote the electron (hole)
creation operators are associated with free charge carriers.
The transition from $|i\>$ to $|f\>$ is enabled by  electrostatic interactions $V_{mn}$
between molecules $m$ and $n$ with a transition matrix of the form
\bea
\label{transanni}
\<f| \hat{V}|i\> &=& 2 N^{-1} \sum_{\bf l,n}\exp[{\bf k.(n-l)}] \\ \nonumber 
&\times&
\left[\<{\bf 0},{\bf m};{\bf e+h},{\bf l }\;|
V_{mn}|{{\bf e \atop \vert} \atop {\bf  h}},{\bf n};{{\bf e \atop \vert} \atop {\bf h}},{\bf m}\> 
-\frac{1}{2} \<{\bf e+h},{\bf l};{\bf 0},{\bf n }\;|
V_{mn}|{{\bf e \atop \vert} \atop {\bf  h}},{\bf n};{{\bf e \atop \vert} \atop {\bf h}},{\bf m}\> 
\right ]
\eea
The first term within the square bracket describes 
recombination of a free exciton (denoted by 
${{\bf e \atop \vert} \atop {\bf h}}$) at site {\bf m},
with  the released energy inducing  the 
 exciton at {\bf n} into a pair of free carriers (denoted by {\bf e+h}) at site {\bf l}.
 A similar interpretation holds for the second term, with  
recombination of the  free exciton taking place at site {\bf n} instead.
The initial and final state energies are obtained as
\bea 
\label{conanni}
E_i &=&  W_0 + 2 E_{0}(K) + {\hbar ^2 K^2 \over 2M^\ast} \\ \label{conannif}
E_f &=& W_0 + E_e(K) + {\hbar ^2 K^2 \over 2 m_e^ \ast }
\eea
$W_0$ is the total electronic energy of crystal
with all molecule in their ground states and $E_{0}(K)$ is 
the exciton energy given in Eq. (\ref{exenergy}). 
$E_e(K)$ denotes the free electron energy.  
Using the Golden Rule formula of Eq. (\ref{exphonimp})
as well as Eqs. (\ref{transanni}) and (\ref{conanni}), we obtain 
the exciton-exciton annihilation rate $R_{an}$ as
\be
\label{anrate}
\frac{1}{T_{an}} = R_{an} \approx \frac{10 \; n_{ex}} {\hbar^4}\; \left[ {m_e^\ast}^{3/2} \;E_T^{5/2} 
\; {N^\ast}^6 \; {d_0}^6 \right ]
\ee
where  $n_{ex}$ is the concentration of  free excitons and 
$N^\ast$ and $d_0$ are defined below Eq. (\ref{formrate}).
 $E_T$,  the transport gap is used to estimate the square of the
transition matrix in Eq. (\ref{transanni}). This approximation
is justified as  Eq. (\ref{transanni}) describes
transition associated with the  
energy required for the formation of a free electron and
hole pair as quantified by the transport gap. 
  
We estimate the exciton-exciton annihilation rate $R_{an}$
for the naphthalene  host crystals (${\rm C_{10}H_8}$) using
$E_T = 3 \times 10^4 {\rm cm}^{-1}$, $n_{ex} \sim 10^{12} {\rm cm}^{-3}$,
$\xi=1283$ kg m$^{-3}$, $N \times d_0= 7.8 \times 10^{-10}$ m,
and obtain  $T_{an}=\frac{1}{R_{an}} \approx 10^{-11}$s. Thus 
exciton-exciton annihilation time $T_{an}$ is comparable to the 
Grover-like search time, $T_N \approx 100$ps at high
concentration of  free excitons. 
At even higher exciton concentration $n_{ex} >  10^{15} {\rm cm}^{-3}$,
this process may dominate the exciton dynamics leading to a breakdown
of the  Grover-like search process. Hence it is imperative that  
the exciton concentration is set low so that  exciton-exciton annihilation
is not disruptive to the  search process, in 
naphthalene-doped organic  crystals. We note the crucial dependence
of $R_{an}$ on   ${N^\ast}$, which yields a measure of packing density of the
searched sites. Hence the distribution of searched sites plays an important
role in the influence of exciton-exciton annihilation, which is expected
intuitively as well.

\subsection{\label{errorff}Exciton fusion and fission.}

During exciton fusion,  two triplet excitons 
combine to form a singlet exciton as shown below
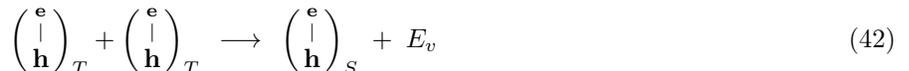
\be
\left ( {{\bf  e \atop \vert} \atop {\bf h}} \right)_T 
+ \left ( {{\bf  e \atop \vert} \atop {\bf h}} \right)_T
\;\;{\longrightarrow}
 \;\;\left ( {{\bf  e \atop \vert} \atop {\bf h}} \right)_S \; + \; E_v
\ee
The subscripts $T$ and $S$ correspond to the triplet and singlet exciton respectively,
with  excess energy $E_v$  generally released as molecular vibrations. 
In naphthalene organic system, the energy difference  between 
two triplet excitons and a singlet exciton is given by
$ 2 E(\sigma_S=1)- E((\sigma_S=0) =E_v$= 1.3 eV, with
the  triplet exciton energy tending to 
decrease faster than the  singlet exciton \cite{Jundt}.
A reverse process in which a singlet exciton undergoes 
conversion into a pair of triplet excitons is termed
exciton fission. Hence in the fusion (fission) process, 
a  singlet (triplet) exciton  is formed with 
all excess energy is released (absorbed) via intra- or inter-
molecular vibrations. 
The calculation of rates of  exciton fission or fusion
is similar to that for exciton-exciton annihilation for which
free charge carriers were  created in the final state (see Eq. (\ref{conannif})).
 We therefore do not give  quantitative details  
of the process, however we note  the sensitive dependence of such processes
on $n_{ex}$, the concentration of  free excitons as obtained in 
Eq. (\ref{anrate}). This was also  noted in a recent  work 
 with rates estimated
in the picosecond range $\sim 1-10$ ps  for closely related 
naphthalene-related systems \cite{excanni}. We expect that
 at low exciton concentration and  packing density of the
searched sites ${N^\ast}$,  exciton fusion and 
fission  processes are less likely to disrupt the overall dynamics
involved in the Grover's like search process. 

\section{\label{zeno} Exciton dynamics control via the Zeno effect.}

Quantitative estimates for the  naphthalene doped organic systems
show that several  mechanisms, in particular downward relaxation due to 
exciton-phonon interaction, exciton-exciton annihilation, exciton fusion
and fission process become dominant at 
raised temperatures and high exciton concentration which translates
as packing density of the searched sites. 
Overall we note that the  coherent term $T_0$ in  Eq. (\ref{stimef2}) 
 associated with the maximum bandwidth 
 still remains the dominant pathway to energy transfer in comparison
to  the Grover's search term $T_N$ (Eq. (\ref{stimef3})),
relaxation term $T_{\rm scat}$ (Eq. (\ref{scatTime})), formation time
 $T_f$ (Eq. (\ref{formrate})) and  exciton-exciton annihilation
time,  $T_{\rm an}$ (Eq. (\ref{anrate})). Nevertheless, the exciton
dynamics associated with Grover's search term $T_N$ 
can be carefully harnessed via 
the quantum Zeno effect \cite{Sudar}. This effect
 embodies the controlling effect of a measurement
process  whereby a system monitored to determine whether it remains in a
particular state resists making transitions to alternate states.
In recent years, this  idea has been
studied using an adiabatic theorem \cite{facchi,facchi2} in which different outcomes are eliminated
and the system evolves as a group of exclusive quantum Zeno subspaces within the
total Hilbert space. The quantum Zeno effect  can be utilized
to exclusively direct the Grover-like search process to proceed uninterrupted while
suppressing  alternate mechanisms of exciton transport and dynamics.
Such directed control of many-body dynamics via  Zeno effect has been experimentally
demonstrated in recent works \cite{yam,lau}, with potential 
applications in quantum information systems. We are unaware of the use
of Zeno effect to switch selected interactions ``on and off" in molecular systems.

\section{\label{cycle} Search on a fluctuating long-range interacting cycle (LRIC) graph lattice}

 A long-range interacting cycle (LRIC) graph is an extension of the 
nearest-neighbor model and a simpler variant of the model considered earlier in 
 Eq. (\ref{lrange}). In a LRIC graph lattice, 
all two nodes of distance $m$ are connected \cite{xu}, a key feature which ensures that 
all lattice sites have the same connectivity of 4.
These newly added edges serve as shortcuts in the cycle graph, and serve to increase
the cross paths and hence  bandwidth $B$ of the lattice system. 
The  one-dimensional LRIC graph lattice is denoted
by two parameters, $N$ and $m$ and the exciton Hamiltonian present in such a 
 lattice appear as
\be
\label{exLRIC}
\hat{H}_{ex}= \sum_{l} E_l B_{l}^\dagger B_{l}+
J \sum_{l}  B_{l}^\dagger B_{l+1}+
J \sum_{l} B_{l}^\dagger B_{l-1}+
J \sum_{l} B_{l}^\dagger B_{l+m}+
J \sum_{l} B_{l}^\dagger B_{l-m}
\ee
where $E_l$ denotes the onsite energy and $J>0$, and the intersite 
energy is assumed to be independent of
the distance between any two sites. 
Using the Fourier transform in Eq. (\ref{exF}), the eigenenergies
are  obtained as $E_{n,m}=\Delta E+2 J( \cos K + \cos m K)$
with momenta  $K= \frac{2 \pi k}{N}$ within
the Brillouin zone $k \in [1,N]$. The  LRIC  Hamiltonian 
is diagonal in $K$ space
\be
\label{LRICen}
\hat{H}_{ex} = \sum_{K} E_{n,m}(K) B_{ K}^\dagger B_{K} 
\ee
The bandwidth $B$ of a Frenkel  exciton in a LRIC lattice therefore has
a simple form $B=2 J \cos K +2 \cos m K$. We now consider that the lattice
points of a LRIC graph undergo small vibrations, and using  the Green's function 
approach considered in the earlier Section, we can examine the effect of 
$m$ on the search time. 
We  consider  a large $N$ so that the impurity trap falls within the 
bandwidth $B$. The exciton bandwidth is also considered to include
the acoustic phonon bandwidth. At small $m$, 
$B(K \rightarrow 0, N \rightarrow \infty) \approx 2 J (1-(\frac{2 \pi m }{N})^2)$
and  using  Eq.(\ref{stime}), we obtain  a search time component $T_N$ that  scales
as $N^2$. However for large $m$, the bandwidth oscillates strongly  depending  on $m$.
Using  $N = p m$ where integer $p$  provides a measure of the lattice network extension,
and considering small values of $\cos m K$, we obtain  
\bea
\label{stimeLR}
\frac{1}{T_s}  &\approx& \frac{1}{T_0}- \frac{1}{T_N} 
\\ \nonumber
T_p &\approx& \frac{T_0}{\cos(\frac{2 \pi}{p})}
 \eea
where as in  Eq.(\ref{stimef}), $T_0$ is an 
$N$-independent  term. The search time component $T_p$ decreases 
gradually with  $p$ so that  search time $T_s$ increases
overall as is expected with an increasing lattice network extension. 
We note that $T_s$ in Eq. (\ref{stimeLR}) does not resemble
the Grover-like form obtain earlier in Eq. (\ref{stime}), due to the
vast reduction in exciton delocalization property associated with the simpler
exciton Hamiltonian structure of the LRIC graph lattice.
Nevertheless the explicit dependence of  search time $T_s$ on $p$ illustrates
the importance of the efficiency of excitation transfer in quantum searches.
The search time $T_s$ is optimum for large $m$, which agrees well with 
the result that the exciton survival probability decays faster on networks of
large $m$ \cite{xu}.

\section{\label{con} Discussion and Conclusion}
 We  have demonstrated the plausibility of  a  Grover-like 
search problem  by means of Frenkel exciton trapping process at
a shallow isotopic impurity. 
We have considered several 
 mechanisms which may compete on equal terms
with our proposed  Grover's search mechanism.
Quantitative estimates for the  naphthalene doped organic systems show that
 Grover's search mechanism remains  plausible under suitable conditions and may
be physically implemented with assistance  from  Zeno effect. 
We emphasize the striking resemblance of a derived explicit 
term for exciton trapping time to the 
  well-known Grover's search time, which  includes  notable speedup features
in the characteristic  range $1 < \mu < \frac{3}{2}$ for a 
graph lattice with  long-range intersite interactions. These 
  parameters  provide  optimum conditions required to trap the exciton in 
the fastest time. We suggest use of Zeno effect associated with
experimental  measurements as a viable means of
eliminating  ubiquitous disruptive interactions 
and harnessing the search process,
with  emission of fluorescence signifying the  
 end of a quantum search.

A recent work on light-harvesting networks \cite{gaab} showed the role of network size and connectivity
as critical factors which lead to optimum energy transfer in graph
lattice systems, and our results agree well with this observation.
We emphasize the  significance  of a derived connectivity dependent
   time component $T_0$,  associated with  intermolecular cooperativity and
 coherence  properties of the entangled exciton  as a  dominant
source of excitation transfer  mechanism. The derived explicit term
can be utilized to examine  high efficiencies of  energy transfer in  photosynthetic systems \cite{alex}. 
Estimates of  various search times and excitonic mechanisms obtained in this work 
can be verified  by experiments employing fast  femtosecond  time resolution
spectroscopic techniques \cite{expt1}.
Finally, our results show that  the natural evolution mechanism of
the Dyson propagator within the Green's function formalism may play a
 critical role in the construction of efficient
quantum search devices, and in 
bridging the gap between excitonic theory and its use in quantum information processing.

\bigskip

\section{Acknowledgments}
I gratefully acknowledge  useful comments from the anonymous referee.

\end{document}